# Localization of Two-dimensional Electron Gas in $LaAlO_3$/$SrTiO_3$ Heterostructures

T. Hernandez[a], C.W. Bark[b], D.A. Felker[a], C. B. Eom[b], M.S. Rzchowski[a]

[a]Department of Physics, University of Wisconsin, Madison, WI 53706, USA
[b]Department of Materials Science and Engineering, University of Wisconsin, Madison, WI 53706, USA

## Abstract

We report strong localization of 2D electron gas in $LaAlO_3$ / $SrTiO_3$ epitaxial thin-film heterostructures grown on $(LaAlO_3)_{0.3}$-$(Sr_2AlTaO_3)_{0.7}$ substrates by using pulsed laser deposition with *in-situ* reflection high-energy electron diffraction. Using longitudinal and transverse magnetotransport measurements, we have determined that disorder at the interface influences the conduction behavior, and that increasing the carrier concentration by growing at lower oxygen partial pressure changes the conduction from strongly localized at low carrier concentration to metallic at higher carrier concentration, with indications of weak localization. We interpret this behavior in terms of a changing occupation of Ti $3d$ bands near the interface, each with a different spatial extent and susceptibility to localization by disorder, and differences in carrier confinement due to misfit strain and point defects.

Since the discovery of a conducting two-dimensional electron gas (2DEG) at the interface between the band insulators LaAlO$_3$ (LAO) and SrTiO$_3$ (STO) [1], multiple efforts have been made to understand the transport mechanism and improve the observed mobilities [2--9]. Temperature dependent transport properties have found low-temperature mobilities broadly ranging from $10^1$ to $10^4$ cm$^2$/V s [1,6,7,9]. Varying transport properties in LAO/STO have been observed under different growth conditions, in particular varying oxygen partial pressures during the LAO layer deposition [6,7]. This suggests that oxygen vacancies play a role under some growth conditions, consistent with the known influence of oxygen vacancies in STO [10,11], and that growth conditions can be optimized to control the mobility and concentration of interfacial carriers.

Measurements of the 2DEG spatial extent indicate a dense sheet of carriers, 2 – 4 nm thick, decaying into the STO layer [12--14]. Carriers occupy the STO Ti 3$d$ bands near the interface [12,15], split by crystal field and structural distortions [15]. In particular, the Ti 3$d$ $t_{2g}$ sub-band is split into a low energy $d_{xy}$ singlet, and a higher energy $d_{xz}$ and $d_{yz}$ doublet [8,16]. The lowest conduction band of Ti 3$d$ states on the first TiO$_2$ layer has strong two-dimensional (2D) character, and hence susceptible to localization due to disorder. Carriers in the higher-energy $d_{xz}$ and $d_{yz}$ bands have large effective mass along the plane, and are also somewhat prone to localization. Carriers occupying $d_{xy}$ subbands spread over several TiO$_2$ layers away from the interface and so are less susceptible to localization by interfacial disorder [8]. These 3D, non-localized subbands are occupied when interfacial carrier concentrations are above $10^{14}$ cm$^{-2}$ [16].

A complete understanding of oxide 2DEGs requires control of both the LAO and STO layer properties, through growth of all-thin-film LAO/STO heterostructures on bulk substrates. Substrate choice provides the opportunity to manipulate the interfacial 2DEG. Electron transport at the LAO/STO interface has been reported for heterostructures grown on Si [17], LSAT [18,19] and NdGaO$_3$ (NGO) [18]. This approach can lead to several effects, including increased interfacial disorder when growing on an additional interface, strain effects introduced by substrate misfit, a different interfacial structure, and different Ti 3$d$ band occupations.

Effects of disorder on transport in the LAO/STO interfacial system have been reported previously. Charged dislocation cores were suggested to be responsible for increased resistance and lower mobilities [20], and weak localization effects were identified through magnetotransport measurements[21]. A reduction of mobility in field effect experiments was attributed to confinement of carriers closer to the disordered interface, increasing scattering [22]. A metal-insulator transition tuned by field effect was suggested to arise from disorder-induced localization or strong Coulomb interactions [23].

Misfit strain in the STO layer grown on different substrates can change the crystallographic phase, and band structure and occupation. Strain in thin film STO has been previously identified to increase mobility by 300% [24], attributed to band structure changes, and carrier concentration changes in LAO/thin-film STO structures were attributed to a strain-induced electric polarization [18]. In addition, theory predicts that complex structural phases appear in strained thin film STO[25], and indications of these phases have been observed [26].

We present a magnetotransport study of epitaxial thin-film LAO/STO on (LaAlO$_3$)$_{0.3}$-(Sr$_2$AlTaO$_3$)$_{0.7}$ (LSAT) substrates, demonstrating strongly localized transport at low temperatures on structures with LAO layers grown at high (10$^{-3}$ mbar) oxygen partial pressures ($p_{O_2}$). Increased carrier concentration and metallic behavior is observed in samples where the LAO layer was grown at lower $p_{O_2}$. This leads to the occupation of subbands in Ti layers further from the LAO/STO interface that are less susceptible to the strong interfacial disorder. We support this interpretation with low temperature magnetoresistance measurements.

Fully coherent LAO/STO thin film heterostructures were grown on LSAT substrates at different oxygen partial pressures using PLD with *in situ* high-pressure reflection high-energy electron diffraction (RHEED) as described previously [18]. LAO overlayers were grown with $p_{O_2}$ of 10$^{-3}$, 10$^{-4}$ and 10$^{-6}$ mbar, and thicknesses of 15 and 20 uc, were grown on 50 uc STO. We have found the crystalline quality of the SrTiO$_3$ template, determined by the FWHM of the x-ray diffraction rocking curve, to be far superior to bulk single crystal (001) SrTiO$_3$ substrates. Aluminum wires bonded directly at the sample corners in a four-point Van der Pauw geometry allowed the determination of sheet resistance ($R_S$), Hall coefficient ($R_H$) and magnetoresistance (MR = $(R_S(H) - R_S(0))//R_S(0)$). We used magnetic fields of up to 8.3 T perpendicular to the film surface at temperatures between 3 and 300 K. The Hall voltage was linear in magnetic field. Sheet carrier concentrations ($n_s$) and Hall mobilities ($\mu_H$) were calculated according to $n_S = -1/(e\, R_H)$ and $\mu_H = (e\, n_S\, R_S)^{-1}$, where *e* is the electron charge.

The samples grown at 10$^{-3}$ mbar (high-pressure samples) with LAO thicknesses of 20 and 15 uc showed metallic behavior at high temperatures and an upturn in R$_S$ near 100 K, reaching values as high as $5\times10^5$ Ω/☐ at 3 K, as seen in Fig. 1. These low temperature R$_S$ were much higher than the quantum of resistance $h/e^2 = 25.8$ kΩ, suggesting strong localization. The temperature dependence of R$_S$ is consistent with 2D Mott variable range hopping (VRH) $\rho = \rho_0 \exp[(T_0/T)^{1/3}]$ [27], (dashed lines in Fig. 1). It was not consistent with Efros-Shklovskii VRH, suggesting that Coulomb interactions do not play a major role [27]. Both thicknesses of LAO grown at this $p_{O_2}$ show similar behavior.

This high-pressure sample displayed a large negative MR below 15 K that increased with decreasing temperature and increasing magnetic field, becoming more than -25% at 3 K and 8.3 T (Fig. 2a). The MR sample did not saturate with decreasing temperature (inset to Fig. 2a) and did not follow Kohler scaling (Fig. 2b) [28]. This indicates that the classical orbital MR observed for out-of-plane field in high mobility LAO/single crystal STO 2DEGs [29,30] is not dominant. Similar but less pronounced negative out-of-plane MR has been observed in LAO/STO structures with high zero field $R_S$ (28.5 kΩ/☐) at 4.2 K [5]. *In-plane* negative MR attributed to magnetic scattering has been reported in high-mobility samples[31], but with conventional positive out-of-plane MR. The large negative out-of-plane MR, lack of Kohler scaling, high zero field $R_S$, and VRH temperature-dependence of $R_S$ all indicate strong localization [32] of interfacial carriers in the high-pressure grown sample.

The 10$^{-4}$ and 10$^{-6}$ mbar (low-pressure) grown samples, both with 20 uc LAO thicknesses, showed transport properties qualitatively different from the high-pressure samples. We observed metallic behavior down to 75 and 50 K respectively (Fig. 1), with small up-turns down to 3 K. The low-pressure samples displayed small *positive* MR with maxima near 7 T (Fig. 3a). The positive MR of the high-pressure samples at 8.3 T also increased with decreasing temperature without apparent saturation, as seen in the inset of Fig. 3a. This MR does not follow Kohler scaling (shown in Fig. 3b for the 10$^{-6}$ mbar sample, the behavior of the 10$^{-4}$ mbar sample is similar), and the downturn in MR near 7 T is consistent with weak localization [32].

We attribute the different carrier concentrations and transport behavior of samples grown at oxygen growth pressures to a different contribution of oxygen vacancies. The samples grown at $p_{O_2}$ = 10$^{-4}$ and 10$^{-6}$ mbar showed room temperature $n_S = 2 - 5 \times 10^{14}$ cm$^{-2}$ (Fig. 4a), higher than the polarization catastrophe prediction of half electron per unit cell or $3.3 \times 10^{14}$ cm$^{-2}$, suggesting, that even here some carriers arise from oxygen vacancies. The sample grown at 10$^{-3}$ mbar showed a room-temperature $n_S = 6.2 \times 10^{13}$ cm$^{-2}$, lower than both the polarization catastrophe prediction and the calculated critical density at which carriers begin to occupy 3*d* subbands [16]. As a check, we found that 100nm thick, oxygen deficient STO thin films (low pressure grown) on LSAT substrates without an LAO overlayer were insulating, as also reported in the literature [10], indicating that an LAO/STO interface is required in order to show conductivity in oxygen deficient STO on LSAT.

The low-pressure samples displayed room temperature mobilities of order $10^0 cm^2$/V s, and of order $10^1 cm^2$/V s at 3 K (Fig. 4b). We fit the scattering time in a single band model to $\tau = (\tau_1^{-1} + \tau_2(T)^{-1})^{-1}$, including a temperature-independent impurity scattering time $\tau_1$, and a temperature-dependent $\tau_2(T) \sim (T/\bar{T})^\alpha$ attributed to lattice interactions. We find $\alpha \sim 2.1$ and $\alpha \sim 2.4$ for the 10-6 and 10-4 mbar samples, respectively, in general agreement with exponents of $2 - 2.7$, reported for LAO on single crystal STO heterostructures [9,29,34]. Weak-localization effects are destroyed by inelastic scattering in this high-temperature regime [35]. At low temperatures, the mobilities saturate at values two orders of magnitude smaller than the highest observed at the LAO/single crystal STO interface [0,3,6,7]; but are consistent with recently reported results obtained in a LAO/STO on LSAT heterostructure [19].

We argue that two factors dominantly contribute to the low-temperature mobility of carriers at the LAO/STO interface: a change in spatial extent of carriers at the interface and disorder at the interface. Varying mobilities in field effect experiments have been discussed in terms of a changing spatial confinement of the electrons with applied electric field, with carriers closer to the interface more susceptible to scattering and localization by interfacial disorder [22,23]. LAO grown at lower pressures leads to the occupation of bands further into the STO layer, which are less susceptible to interfacial localization.

Strain in the STO layer may also affect the interfacial mobility. Compressively strained STO has been reported to undergo transitions to more complex structural phases than the tetragonal phase observed below 110K in unstrained STO [25,26].

In addition, a symmetry lowering transition has been reported at 150K in LSAT, with neutron diffraction strongly suggesting a tetragonal distortion [36]. These may lead to a more complex LAO/STO interface structure than that of unstrained STO heterostructures, and possibly to higher levels of disorder, in addition to the disorder discussed previously. It has also been reported that compressively strain in fully coherent LAO/STO heterostructures on LSAT produces a polarization that points away from the LAO/STO interface, reducing the carrier concentration by the field-effect [18].

Our measured carrier concentrations indicate an introduction of more carriers in the low-pressure samples, which will begin to fill subbands distributed in $TiO_2$ layers deeper in the STO film that are less susceptible to interfacial disorder [8]. The carrier densities in the low-pressure samples are above the $\sim 10^{14} cm^{-2}$ carrier concentration at which $3d_{xz}$ and $3d_{yz}$ subbands in deeper Ti layers begin to be occupied [16]. These carriers will begin to dominate electrical transport measurements at low temperatures, when the carriers in lower energy bands confined closer to the interface, having greater 2D character, become highly localized, as suggested by the low temperature $R_S$ in the high-pressure sample. The weakly localized character of the MR in the low-pressure samples further supports the hypothesis that the carriers in these samples are less prone to interfacial disorder localization. Disorder at the interface appears to be greater in LAO/STO on LSAT substrates than in the higher-mobility LAO on single crystal STO heterostructures. Another possible cause of the greater disorder is the introduction of more point defects during deposition growth at two interfaces, as opposed to one.

It is the increased disorder in these heterostructures that shows the dramatic difference in transport properties at different carrier concentrations.

We have shown that the 2DEG LAO/thin film STO interface on LSAT substrates grown at high $p_{O_2}$ shows strongly localized transport at low temperatures, supported by both the temperature dependence of $R_S$ and the MR. The low-pressure grown samples displayed greater sheet carrier densities and weakly localized behavior at low temperatures. We have argued that the additional carriers in the low-pressure samples begin to occupy subbands in Ti layers further from the LAO/STO interface that are less susceptible to interfacial disorder.

Additional work is necessary to fully understand the source of higher disorder and lower mobility at the LAO/STO heterointerface on LSAT, but our measurements suggest contributions from both intrinsic factors due to strain, and extrinsic factors due to defect control. The relative importance of these has not been determined, but advances in interface control during growth may substantially improve the mobility. We also expect that electric dipole moments in the complex phases of strained STO affect transport and carrier confinement at the interface.

Acknowledgments

This work was supported by NSF grant 0906443, FRG: Switchable Two-Dimensional Materials at Oxide Hetero-Interfaces.

**Figure Captions**

FIG. 1. Temperature dependence of the sheet resistance. Two high-pressure ($10^{-3}$ mbar) grown samples with 20 and 15 uc LAO thicknesses (triangles and diamonds respectively), and two low pressure samples with LAO thickness of 20 u.c. grown at $10^{-4}$ and $10^{-6}$ mbar grown samples (squares and circles respectively). The quantum of resistance $h/e^2 = 25.8$ kΩ is indicated as a dotted line. Dashed lines show fits for the two dimensional Mott type variable range hopping temperature dependence for both $10^{-3}$ mbar samples on the range 3 to 100 K. Solid lines for the $10^{-4}$ and $10^{-6}$ mbar samples are guides to the eye.

FIG. 2. Perpendicular field magnetoresistance for the high-pressure sample. (a) MR behavior of the $10^{-3}$ mbar sample at T = 3.0 K (circles), 5.3 K (squares) and 10.0 K (diamonds). The inset shows the temperature dependence of the MR at 8.3 T. (b) Kohler scaling of the MR for the same sample at the same temperatures.

FIG. 3. Perpendicular field magnetoresistance for the low-pressure samples. (a) MR behavior of the $10^{-4}$ (squares) and $10^{-6}$ (circles) mbar grown samples at T = 3.0 K. The inset shows the MR of the temperature dependence of the $10^{-6}$ mbar sample at 8.3 T. (b) Kohler scaling of the MR for the $10^{-6}$ mbar grown sample at T = 3.0 K (forward triangles), 5.3 K (backward triangles) and 10.0 K (diamonds).

FIG. 4. Temperature dependence of (a) the sheet carrier density and b) the Hall mobility. The $3.3\times10^{14}$ cm$^{-2}$ sheet carrier concentration predicted by the polarization catastrophe scenario is indicated by the horizontal line. The dashed lines are fits to the mobility for the low-pressure samples, including parallel contributions of temperature independent and power-law temperature dependent of scattering times. All solid lines are guides to the eye.

**Figure 1**

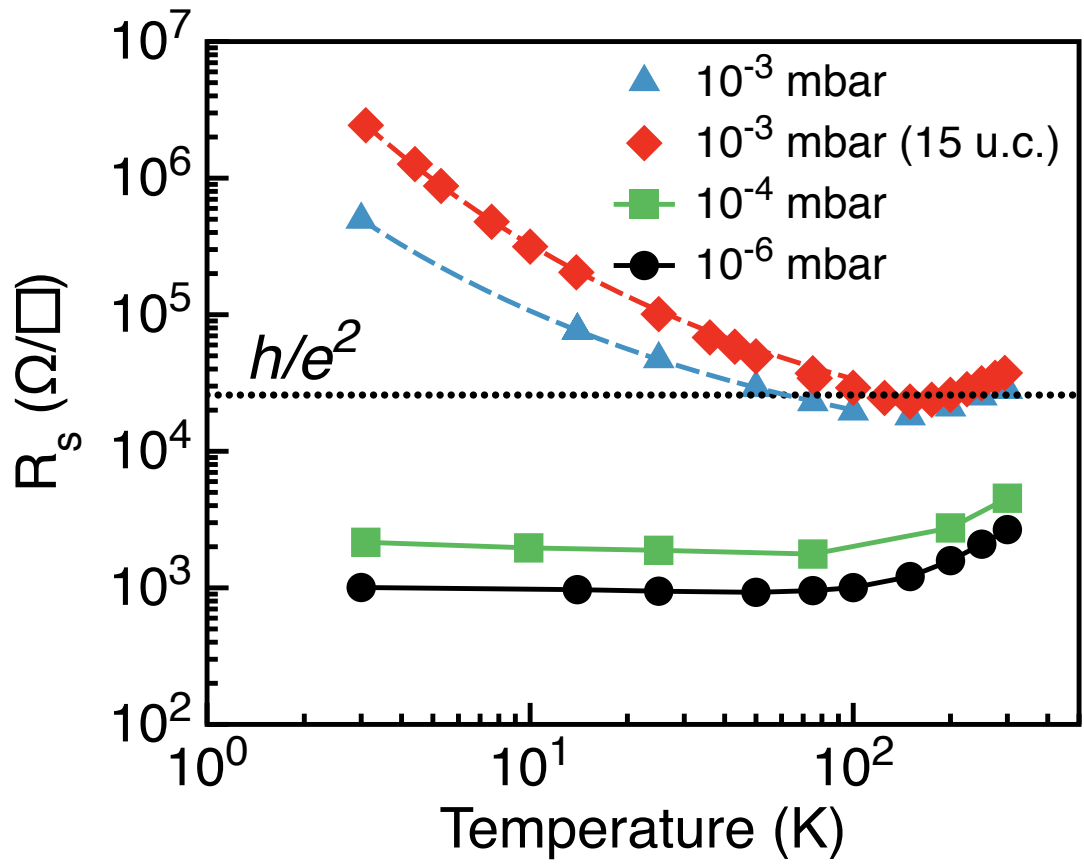

Figure 2

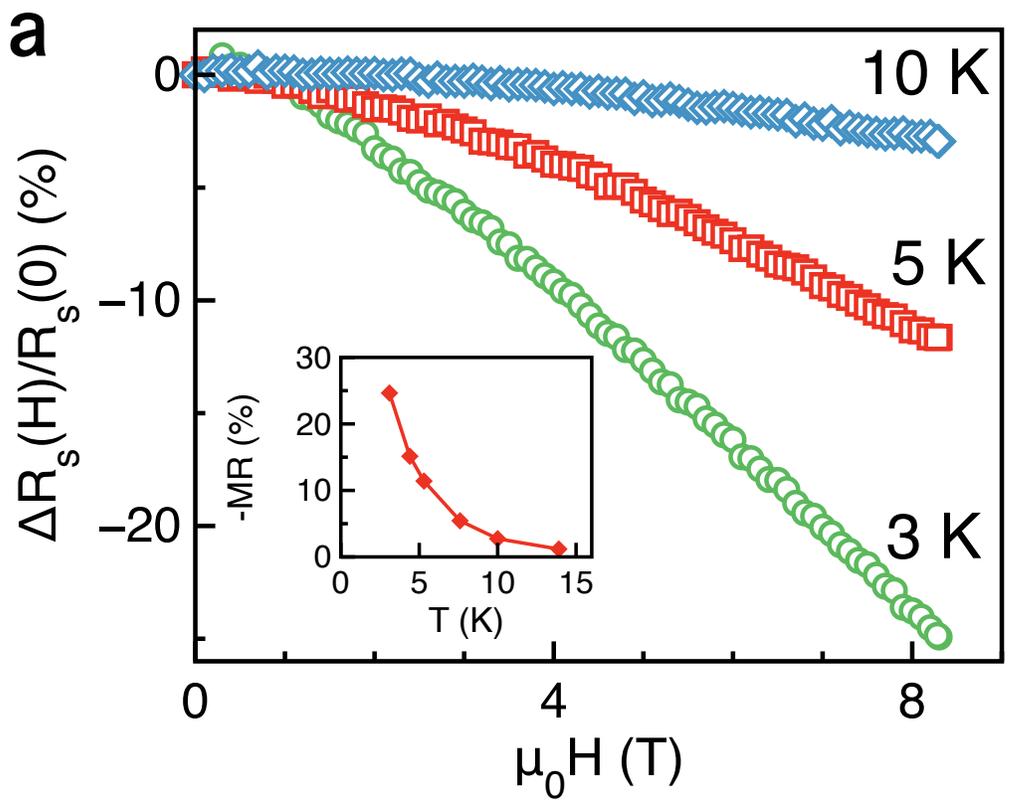

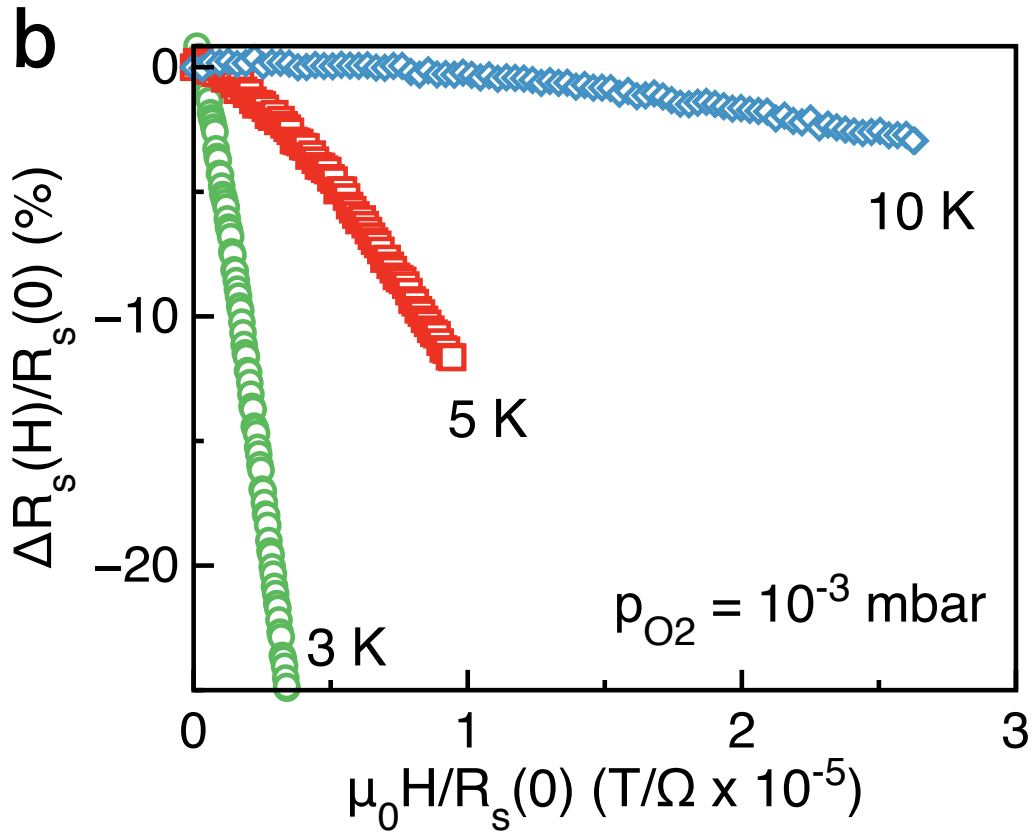

Figure 3

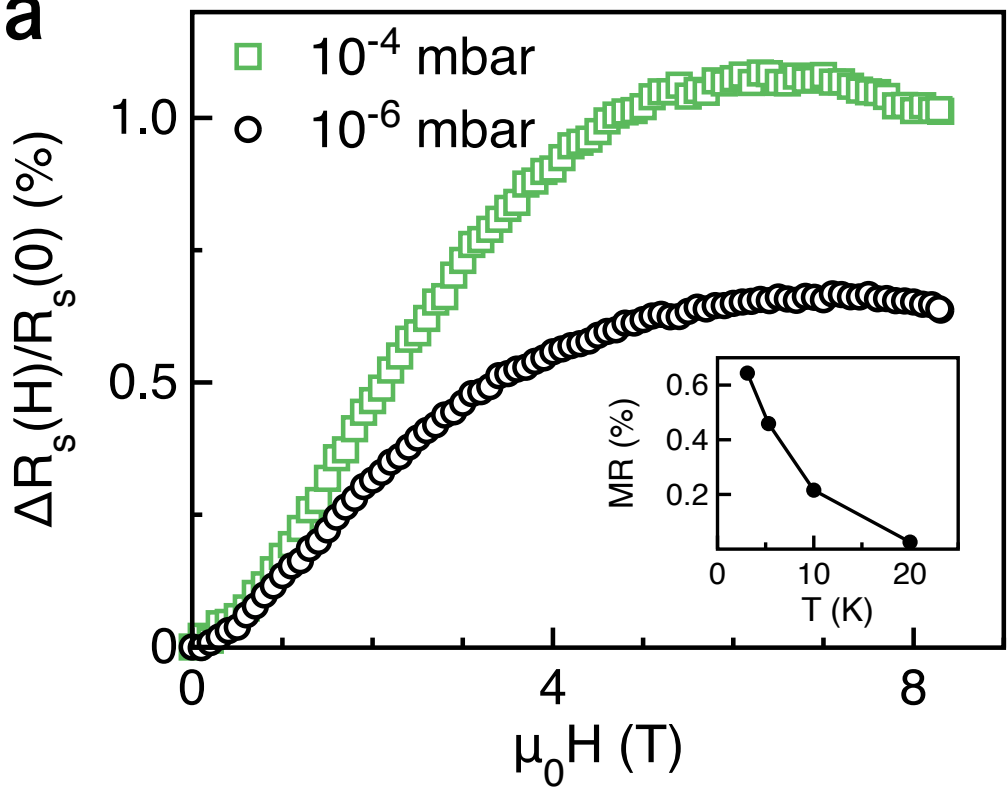

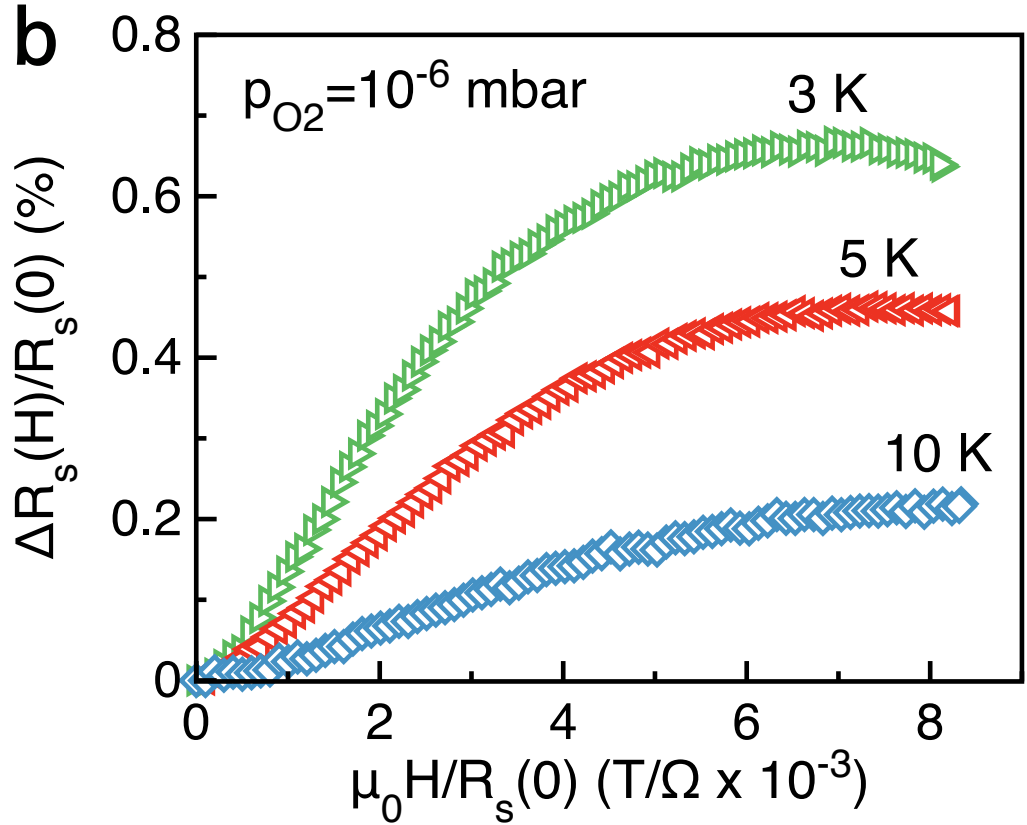

**Figure 4**

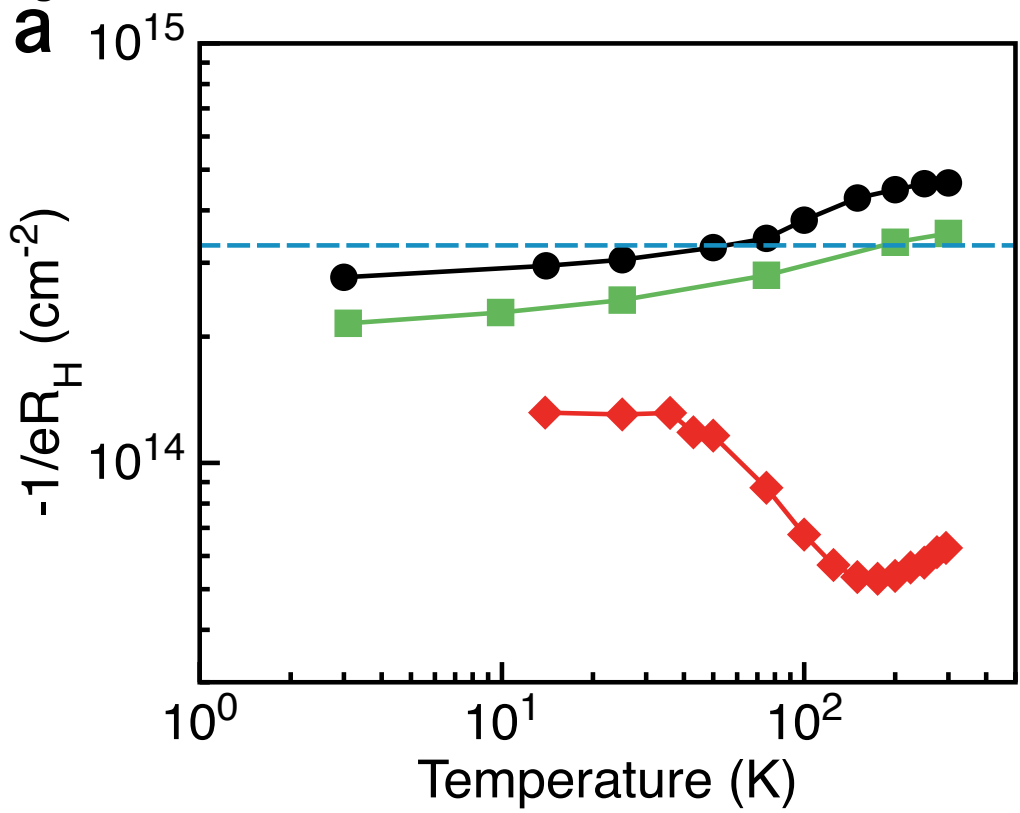

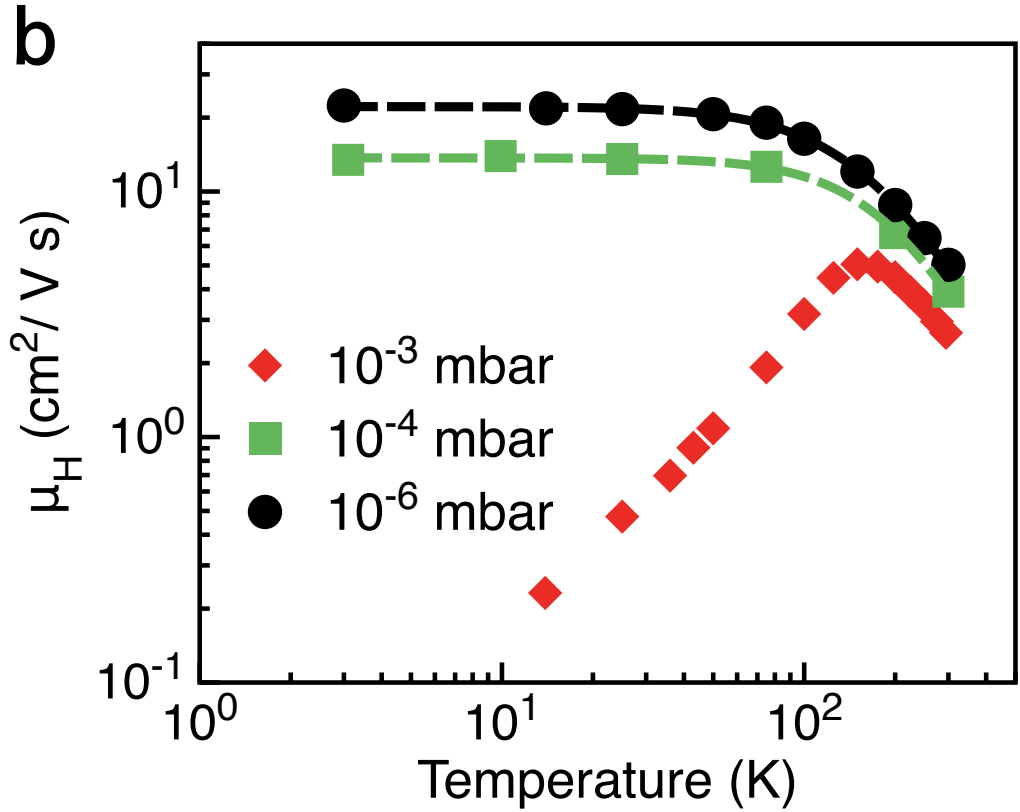